\documentclass[doublecol]{epl2}

\usepackage{amsfonts,amsmath,amssymb}

\definecolor{g-blue}{rgb}{0.83,0.95,1}
\definecolor{g-yellow}{rgb}{1,1,0.7}
\definecolor{g-green}{rgb}{0.9,1,0.9}
\definecolor{green}{rgb}{0,0.6,0}
\definecolor{cyan}{rgb}{0,0.7,0.7}
\definecolor{black}{rgb}{0,0,0}
\definecolor{grey}{rgb}{0.4 ,0.4 ,0.4 }

\def\blue#1{\textcolor{blue}{#1}}

\def \ed {\end{document}}
\def\Fbox#1{\vskip1ex\hbox to 8.5cm{\hfil\fboxsep0.3cm\fbox{%
  \parbox{8.0cm}{#1}}\hfil}\vskip1ex\noindent}  

\def\be{\begin{equation}}\def\ee{\end{equation}}
\def\bea{\begin{eqnarray}}\def\eea{\end{eqnarray}}
\def\bse{\begin{subequations}}\def\ese{\end{subequations}}
\newcommand{\BE}[1]{\begin{equation}\label{#1}}
\newcommand{\BEA}[1]{\begin{eqnarray}\label{#1}}
\newcommand{\BSE}[1]{\begin{subequations}\label{#1}}

\let \= \equiv \let\*\cdot \let\~\widetilde \let\^\widehat \let\-\overline

  \def\1{\bm1} 

\def\<{\left\langle}    \def\>{\right\rangle}
\def\({\left(}          \def\){\right)}
 \def \[ {\left [} \def \] {\right ]}

\newcommand{\B}[1]{{\bm{#1}}}
\newcommand{\C}[1]{{\mathcal{#1}}}    
\newcommand{\F}[1]{{\mathfrak{#1}}}

\renewcommand{\sb}[1]{_{\text {#1}}}  
\renewcommand{\sp}[1]{^{\text {#1}}}  
\def\Sb#1{_{\scriptscriptstyle\rm{#1}}}

\title{Temperature suppression of Kelvin-wave turbulence in superfluids}

\author{Laurent Bou\'e  \and Victor L'vov    \and Itamar Procaccia }
\shortauthor{Laurent Bou\'e \etal}
\institute{
  \inst{1} Department of Chemical Physics, Weizmann Institute of Science, Rehovot 76100, Israel \\

}

\abstract{Kelvin waves propagating on quantum vortices play a crucial role in the phenomenology of energy dissipation of superfluid turbulence.  Previous theoretical studies have consistently focused on the zero-temperature limit of the statistical physics of Kelvin-wave turbulence. In this letter, we go beyond this athermal limit by introducing a small but finite temperature in the form of non-zero mutual friction dissipative force;  A situation regularly encountered in actual experiments of superfluid turbulence.  In this case we show that there exists a new typical length-scale separating a quasi-inertial range of Kelvin wave turbulence from a far dissipation range.  The letter culminates with analytical predictions for the energy spectrum of the Kelvin-wave turbulence in both of these regimes.}

\begin{document}

\maketitle

\section{Introduction}

Quantum turbulence refers to the study of fluids such as~$^{3}$He and~$^{4}$He that remain liquid even at very low temperatures where quantum effects begin to manifest themselves.  In fact, below a characteristic temperature denoted as $T_{\rm c}$ and $T_{\lambda}$ respectively, these fluids acquire a   superfluid component whose viscosity is zero.  Although~$T_{\lambda}$ is   small under usual conditions on Earth, it has been discovered recently that superfluids can also be found in the core of neutron stars where the temperatures reaches values of the order of~$10^9\,$K due to the extremely high density.  This raises the prospect that this state of matter is much more widespread than previously believed~\cite{neutrons}.  Having an inviscid component, one can expect that superfluids should easily reach a turbulent state. Unlike classical fluids where the size of the turbulent eddies is a statistical property that can vary from the small vortices in a stirred cup of coffee to hurricanes in stormy weather systems, the vortices of turbulent superfluids are tightly constrained by the laws of quantum mechanics: the circulation around a quantum vortex is quantized to integer values of the quantum circulation~$\kappa = 2 \pi \hbar / M$ where~$M$ in $^{4}$He is the mass of the atom and in $^{3}$He is the mass of a pair of atoms.  It has been established experimentally by Vinen~\cite{vinen1} that the typical turbulent steady state consists of a complex tangle of quantum vortices as had been speculated by Feynman~\cite{feynman} and later clarified through the numerical simulations of Schwarz~\cite{schwartz1}.

As was already understood by Tisza~\cite{tisza} and Landau~\cite{landau}, superfluids can be modeled as consisting of two interpenetrating fluids with two different velocities $v_s$ (superfluid component) and $v_n$ (normal component). The relative density of each component, ($\rho_s$ and $\rho_n$ respectively), depends on the temperature of the fluid.  In the limit~$T \to 0$, $\rho_n\to 0$ and the fluid becomes totally inviscid.  This immediately raises the question of how a turbulent steady state can be achieved at all with energy injection at the large scales but no obvious dissipation mechanism at the small scales.  The current understanding of this issue is that the appearance of quantum vortices introduces an additional scale (the intervortex mean distance) denoted by $\ell$. For fluid motions of scales much larger than $\ell$ the quantization of the vorticity is irrelevant, and the turbulent spectrum resembles that of a classical fluid with an energy flux towards smaller scales. When the energy cascade reaches scales comparable to $\ell$ the dynamics of the quantized vortex lines becomes relevant. These vortices continuously reconnect with each other generating in the process Kelvin waves on the vortex lines themselves.  For scales smaller than $\ell$ the interaction between vortex lines can be neglected, and then one can study the spectrum of Kelvin waves on isolated vortices. Kelvin waves on isolated vortices are nonlinear, producing an energy cascade again towards even smaller scales. When the scale of these waves becomes comparable to the core radius of the quantized vortex, it is believed that elementary excitations (different in $^{3}$He and~$^{4}$He) radiate out resulting in a dissipation mechanism.   Due to their crucial role, the properties of these Kelvin waves have been studied intensely recently in the limit of weak wave turbulence and at~$T=0$. In particular, their energy spectrum at $T=0$ was found in~\cite{victor1,victor2}, see eq.~(\ref{LN}) below.

So far all of the theoretical studies of the statistical properties of the Kelvin waves have entirely focused on the limit~$T=0$ and almost nothing is known for finite temperatures.  However as soon as~$T\neq0$ there is a friction between the normal component and the superfluid.  This so-called mutual friction is parametrised by a scalar~$\alpha$ and offers another avenue for small scale energy dissipation via thermal damping.  Having in mind that it is difficult to reach temperatures very close to~0 experimentally, generalizing the spectrum of eq.~(\ref{LN}) for finite temperatures represents an essential step towards the formulation of a theory of superfluid turbulence.

\section{Basic equations} \label{ss:base}
The study of non-linear interactions of Kelvin waves in the framework of the Biot-Savart equation (which determines the velocity field induced by vortex lines) goes back to the pioneering paper of Barenghi, Donnelly and Vinen~\cite{BDV} and has attracted a lot of studies since, see~\cite{KozikScan,Sonin2012,Samuels,KS,LLN} for different examples.   Restricting ourselves to wavelengths $\lambda \ll \ell$, the complex and ever-changing topology of the tangle of quantum vortices can be ignored and we can focus on the behavior a single isolated vortex. In this case further progress can be achieved by presenting the Biot-Savart equation in its Hamiltonian form~\cite{sonin,Svi-95}: $ i \kappa \, \partial w / \partial t= \delta H / \delta w^*$,
where~$w(z,t) = x(z,t)+iy(z,t)$ denotes deflections away from a straight vortex aligned with the $z$ direction, the superscript ``$\,^*\, $" denotes complex conjugation and
\begin{eqnarray}\label{H}
 H = \frac{\kappa^2}{4\pi} \int \!\!\! \frac{1+\Re(w'^*(z_1,t)w'(z_2,t))}{\sqrt{(z_1-z_2)^2+|w(z_1,t)-w(z_2,t)|^2}}  dz_1 dz_2.
\end{eqnarray}
Here the deflection angle is $w'(z,t)=\partial w/\partial z$. For $w'(z,t) \ll 1$ one can then follow a long and cumbersome theoretical procedure~\cite{KS,rudenko,victor1} where the Hamiltonian is expanded in power series, represented into the~$k$ space, $w(z,t)\to w_k(t)\equiv a_k(t)/\sqrt{\kappa}$, and subjected to a weakly non-linear canonical transformation, $b_k=a_k + ...$, in order to obtain finally an effective $(1\Leftrightarrow 3)$-wave Hamiltonian~\footnote{Notice that   $\C H_4$  has  the meaning of the interaction energy density  of Kelvin waves (per unit of vortex length), normalized also by the He-fluid density $\rho$. Dimensionally  $[\C H_4]=\un{cm^4 sec^{-2}}$.}
\begin{subequations}\label{H4}\begin{eqnarray}
&&  \C H_4=\frac 16 \sum_{\B k= \B k_1+ \B k_2+ \B k_3}  [\mathcal{W}_{{\bm k}}^{1,2,3} b_{\B k}^* b_{ \B k_1} b_{\B k_2} b_{\B k_3}+ \mbox{c.c.}]\,, ~~~~\\
&& \mathcal{W}_{{\bm k}}^{1,2,3}  \equiv  \frac {-3\Psi}{4\pi\sqrt{2}}\,
\B k\B k_1 \B k_2 \B k_3\ .
\end{eqnarray}
\end{subequations}
Here  the dimensionless  parameter $\Psi \lesssim   1 $ depends on the driving mechanism of the Kelvin waves turbulence and  it will be considered as an external parameter of the theory.

The statistical description of Kelvin wave turbulence is written in terms of dynamical equations for the Kelvin waves action $n_\B k= \C L \< |b_k|^2\>/(2\pi)$ ($\C L$ is the linear size  of the periodic box) which are a classical limit of the occupation numbers $\F n_\B k$: $n_\B k\to \hbar \, \F n_\B k$ for $\F n_\B k\gg 1$~\cite{victorbook}.  Our starting point in this paper is the kinetic equation for $n_\B k$,  supplemented by a mutual friction dissipation proportional to a temperature dependent dimensionless coefficient $\alpha$~\cite{BDV}
\begin{equation}\label{KEa}
\frac{\partial n_\B k}{\partial t}=\mbox{St}_\B k\{ n_{\B k'}\} - \alpha \, \omega_k n_\B k \,, \quad \omega_k= \Lambda \kappa k^2 \big / ( 4\pi )\,,
\end{equation}
Here $\omega_k$ is the Kelvin wave frequency  of wavenumber $\B k$ and the   parameter $\Lambda$ is given by $\Lambda = \ln (\ell / a)$  corresponding to the ratio between~$\ell$ and the vortex core diameter $a$. In typical experiments involving both~$^3$He and~$^4$He, $\Lambda$ varies between~12 and~15. For  small temperatures, when  $\alpha \ll 1$~\cite{expdata} the mutual friction dissipation $\alpha \, \omega_kn_k $ is still important being compatible with    $\mbox{St}_\B k\{ n_{\B k'}\} \ll n_k \omega_k$  for weakly interacting waves.  Note that the frequency shift induced by mutual friction~\cite{BDV} is only a small correction with respect to the $\omega_k$ presented in eq.~(\ref{KEa}) and can safely be ignored at low temperatures.

 Notice that eq.~\eqref{KEa} ignores thermal fluctuations of Kelvin waves which in thermodynamical equilibrium with temperature $T$ are given by the Rayleigh-Jeans distribution $ n\sb{bath}(k)\simeq T/\rho \omega_k$. Here $\rho$ is the density of liquid helium ($0.145\un{g/cm{^3}}$ for $^4$He)
In the framework of standard approach~\cite{victorbook} and   using chain of relations  $w(z,t)\Rightarrow w_k(t) \Rightarrow a_k(t) \Rightarrow b_k \Rightarrow n_k = n\sb{bath}(k) $  we  found that the mean squared deviation,  caused   be  thermal excitations, $\< |w|^2\>\sb{bath}$  is dominated by waves of length $\lambda \sim \ell$, and get the estimate of the dimensionless ratio
\begin{equation}\label{te}
R\equiv \frac{\langle  |w|^2\rangle\sb{bath}}{\ell^2 }\simeq \frac {  T}{\rho \kappa^2 \Lambda \ell}  \sim  (10^{-7}\div 10^{-8})\ .
\end{equation}
 In this estimate  for~$^4$He we took $T\simeq 1$K and $\ell\simeq 10^{-3}\un{cm}$. Since  $R\ll 1$    the only relevant mode of excitation of the Kelvin waves is via reconnection events between crossing vortex lines.

Finally,   for the collision term in eq.~\eqref{KEa} with the   Hamiltonian~\eqref{H4} was derived  in~\cite{victor1}:
\begin{center} \blue{see eq.~\eqref{KEc}  on the next page}\\ \end{center}
\begin{widetext}
\begin{eqnarray}
&&\hskip - 0.5 cm  \mbox{St}_\B k\{ n_{\B k'}\} \!= \!\frac{\pi}{12} \iiint\!\! \upd {\bm k}_1 \upd {\bm k}_2 \upd {\bm k}_3 \left( \mathcal{W}_{{\bm k}}^{1,2,3} \right)^2 \, \Big[\left( \frac{1}{n_\B k} \!-\! \frac{1}{n_ 1} \!- \!\frac{1}{n_2} \!-\! \frac{1}{n_3} \right) \delta \left( {\bm k} - {\bm k}_1 - {\bm k}_2 - {\bm k}_3  \right) \delta \left( \omega_k - \omega_1 - \omega_2 - \omega_3  \right) \nonumber   \\
&&\hskip - .3 cm -  3  \left( \frac{1}{n_1} - \frac{1}{n_k} - \frac{1}{n_2} - \frac{1}{n_3} \right) \delta \left( {\bm k}_1\! - \!{\bm k}\! - \!{\bm k}_2 \!- \!{\bm k}_3  \Big) \delta \left( \omega_1 - \omega_k - \omega_2 - \omega_3  \right)\right] n_k n_1 n_2 n_3 \,, \quad n_j\equiv n_{\B k_j}\,, \quad \omega_j\equiv \omega_{k_j}\ .
\label{KEc}
\end{eqnarray}
\end{widetext}
The collision term in eq.~(\ref{KEc}) preserves the total energy   of the system.  This allows us to rewrite the kinetic eq.~(\ref{KEa}) as a continuity equation for the energy spectrum $E(k)\equiv 2 \omega_k n_k$. Integrating the equation over $\B k$ annuls the contribution of the collision term. This means that, as usual, there exist a flux of energy down the scales, and one can write a continuity equation of the form~\footnote{
Considering an isotropic vortex tangle with mean intervortex distance $\ell$ one relates
the energy (density per unit of vortex length, normalized also by the density $\rho$) flux $\epsilon$ ($[\epsilon]=\un{cm^{4}sec^{-3}}$) over the 1D vortex line to the bulk 3D fluid energy (density per unite mass) flux $\varepsilon$  ($[\varepsilon]=\un{cm^{2}sec^{-3}}$) with the bridge $\varepsilon = \epsilon / \ell^2$.}
\begin{subequations}\label{EB}
\begin{eqnarray}\label{EBa}
  \frac{d \epsilon(k)}{d k} &=& -\alpha \omega_k E(k)\,, \\ \label{EBb}
\epsilon(k) &=& - 2 \int_{{ k'}<{ k}} \mbox{St}_{{\bm k'}} \, \omega_{ {\bm k'}} d{{\bm k'}}\ .
\end{eqnarray} \end{subequations}
This equation has a stationary scale invariant solution in the zero temperature limit when $\alpha=0$. This solution corresponds to a $k$-independent energy flux $\epsilon_k$ equal to the energy influx   $\epsilon_0$.
It was found in~\cite{victor1} and carefully analyzed in~\cite{victor2}:
\begin{equation}
E_0(k)= C_{_{\rm LN}}\frac{\Lambda\, \kappa \,  \epsilon_0^{1/3}}{\Psi^{2/3} \, k^{5/3}}\,,   \quad C_{_{\rm LN}} \approx 0.304.
\label{LN}
\end{equation}
Discussions of alternative spectra of Kelvin waves~\cite{KS,Sonin2012} can be found in~\cite{LLN} and references therein.

At finite temperatures eq.~(\ref{EBa}) contains terms that destroys the scale-invariant solution, producing a characteristic scale $k_{\star}$.  To find this scale we first use the fact that the collision integral converges~\cite{victor2} and make a dimensional estimate of $d \epsilon(k)/ dk $ as $ - \epsilon(k)/  k $.  By balancing the LHS and the RHS of~eq.(\ref{EBa}) using the form of $E(k)$ at $T=0$, one immediately finds the crossover wave vector~$k_\star$:
\begin{equation}\label{ks}
k_{\star}  =  \frac{(4\pi)^{3/4}\sqrt{\Psi \epsilon_0}}
{\big(\sqrt{\alpha C_{_{\rm LN}}}\Lambda\kappa\big )^{3/2}} \ .
\end{equation}
For $k\ll k_{\star}$ the mutual friction term on the RHS of eq.~(\ref{KEa}) is negligible compared to the collision term, and the zero temperature solution is expected to hold up to some corrections that are computed in the next section. We will refer to this region of scales as the quasi-inertial range of Kelvin wave turbulence. For $k\gg k_{\star}$ this solution no longer holds, and we need to find a new solution in the far-dissipation range as is explained in the last section.

\section{Energy spectrum in quasi-inertial range of Kelvin wave turbulence} \label{ss:dif}
In the quasi-inertial region the energy flux $\epsilon(k)$ varies slowly with $k$. This allows us to estimate $\epsilon(k)$ directly from eq.~(\ref{LN}) by substituting there $\epsilon(k)$ instead of the k-independent $\epsilon_0$. This gives the following \emph{algebraic} approximation for $\epsilon(k)$:
\begin{equation}
\epsilon (k)= \Psi ^2 k^5 E^3(k)\big / (C\Sb {LN}\, \Lambda \kappa )^3\ . \label{neweps}
\end{equation}
In the inertial range of scales this equation is equivalent to the solution~\eqref{LN}. We note that in general the kinetic equation with $\alpha=0$ displays another solution which is the equilibrium state $\epsilon(k)=0$ and $E_k=$Const. We can rewrite an equation that is dimensionally equivalent to eq.~(\ref{neweps}) that contains that solution as well.  This is achieved by replacing $E^3(k)$ by $-\frac 1 5 k d E^3(k)/ d k$, where the numerical factor is chosen to reproduce solution~\eqref{LN} for $\alpha=0$. Now we have the \emph{differential} approximation for the energy flux
\begin{equation}\label{e-dif}
\epsilon(k) = - \frac{k^6 \Psi^2}{5 \left(C_{_{\rm LN}} \Lambda \, \kappa \right)^3} \frac{\partial E^3(k)}{\partial k}\,,
\end{equation}
which is a step forward in comparison with the algebraic approximation~\eqref{neweps}, because it accounts not only for the inertial range solution~\eqref{LN}, but also for the thermodynamical equilibrium solution $E(k)=$Const.
For $k < k_\star$ the mutual friction term is small and only slightly modifies the inertial range solution~(\ref{LN}). Therefore we seek a solution of eqs.~(\ref{EBa}) and~(\ref{e-dif}) in the form:
\begin{subequations}\label{e-sub}
  \begin{eqnarray}\label{e-subA}
E(k) &=& E_0(k)\, f(y) \,, \quad y = \frac{15}{4} \Big( \frac{k}{k_{\star}} \Big)^{4/3}\,,
\end{eqnarray}
in which $f(0)=1$. Note the natural normalization of $k$ by the crossover value~$k_\star$. The exponent and pre-factor in the expression for $y$ was chosen such that the final equations below [cf. eq.~(\ref{univ})] for the function $f(y)$ would appear as simple as possible.    Plugging eqs.~(\ref{e-sub}) into eqs.~(\ref{EBa}) and~(\ref{e-dif}) we get a second  order ordinary differential equation for $f(y)$: $\^ {\C D}f(y)=1$, where 
\begin{eqnarray}\label{univ}
 \^ {\C D}f(y) \equiv  4yf(y) f''(y) + 8 yf'(y)^2-11f(y) f'(y) \ .~ 
\end{eqnarray}
\end{subequations}
All the system parameters have dropped out in this equation, making it \emph{universal}.

We could not find an analytical solution of this equation; Instead, we have devised an iterative technique that provides approximations to the exact solution with exponential convergence using a proper choice of a trial function. Using the boundary condition $f(0)=1$, eq.~(\ref{univ}) dictates: $f'(0)=-1/11$. This suggests a simple form of the initial trial function $f_0(y)= 1 - y/y_0$ with $y_0 = 11$. The parameter $y_0$ can be found in a different way from the global requirement
\begin{equation}\label{glob}
\int_0^{y_{0}} f(y) \upd y=5\ .
\end{equation}
This constraint expresses the fact that in the stationary case the rate of energy injection is equal to the rate of energy dissipation.  Simple algebra shows that this condition is satisfied with the function $f_0(y)$ only if $y_{0} = 10$. Since two different ways to determine $y_0$ resulted in $y_0=10$ or 11 one is encouraged to improve the accuracy by proper polynomial corrections:
 \begin{equation}\label{fn}
f_n(y) = \Big( 1 - \frac{y}{y_{0}} \Big) \sum_{m=0}^n a_m \Big( \frac{y}{y_{0}}\Big)^m\,, \quad a_0=1\ .
\end{equation}
We iterate this equation starting from the zeroth order approximation, $f_0(y)$ with $y_{0}=10$ which was determined from the global requirement~\eqref{glob}. Consider next the $2\sp{nd}$ approximation $f_2(y)$,
 in which  coefficients $a_1$ and~$a_2$ are determined by the requirements
 \begin{equation}\label{bc}
f'(0) = -  1/11 \,, \quad
f'(y_{0})= -  1/\sqrt{8 y_{0}}  \,,
\end{equation}
that directly  follows from eq.~\eqref{univ}, while $y_0$ again from eq.~\eqref{glob}.  Even though these conditions can be inverted in order to exactly determine the coefficients, their precise analytical form is not useful here and we prefer to report their numerical values in Table~\ref{coeffs}.  One can already see the excellent convergence properties of our trial function since the coefficients quickly decrease and the value of~$y_{0}$ is already very close to its naive estimate obtained from the zero-order approximation $f_0(y)$.

\begin{figure}
\onefigure[width=0.85\linewidth]{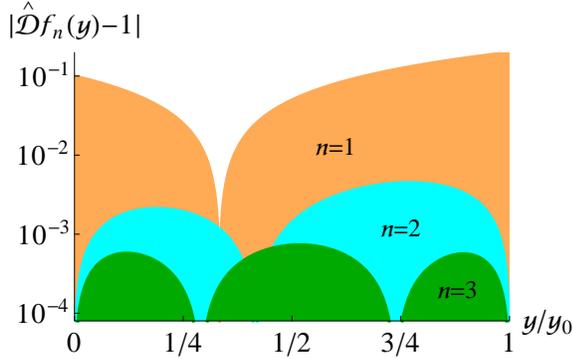}
\caption{\label{f:1} We present a raw comparison of the absolute value of the difference between the LHS and the RHS of~eq.~(\ref{univ}) for different levels of approximation.  The dips correspond to points where both sides of the equations match exactly.}
\end{figure}

\begin{table}
\caption{Value of the coefficients for the various levels of approximations.}
\label{coeffs}
\begin{center}
\begin{tabular}{|c|c|c|c|}
\hline
Approx & $f_0$ & $f_2$ & $f_3$ \\
\hline
\hline
$y_{0}$ & 10 & 9.644 & 9.651 \\
\hline
$a_1$ & 0 & 0.123 & 0.123  \\
\hline
$a_2$ & 0 & -0.025 & -0.035 \\
\hline
$a_3$ & 0 & 0 & 0.011 \\
\hline
\hline
$\chi$ & $10^{-1}$ & $3 \times 10^{-3}$ & $5\times 10^{-4}$ \\
\hline
\end{tabular}
\end{center}
\end{table}

Finally, we move on to our last level approximation $f_3(y)$  in which we keep an additional coefficient~$a_3$ in the expansion of the trial function. In this case, one can express~$a_1$, $a_2$ and~$a_3$ as analytic functions of~$y_{0}$ that is taken now as a free parameter.  Its value is obtained by setting up an optimization problem in which we minimized the mismatch function
\begin{equation}\label{MisErr}
 \chi   =   \Big [ \int_0^{y_{0}}  \big[ \^ {\C D}f(y)   - 1 \big]^2 \upd y \big / y_0\Big ] ^{1/2}\ .
\end{equation}
The numerical values of the coefficients for all three levels of approximation are displayed in Table~\ref{coeffs}.  To illustrate the rapid convergence of the trial functions $f_n$ we present in Fig.~\ref{f:1} plots of $|\^ {\C D} f_n(y)-1|$, the absolute error between the LHS and the RHS of~eq.~(\ref{univ}) for different level of the approximations.

\begin{figure}
\onefigure[width=1\linewidth]{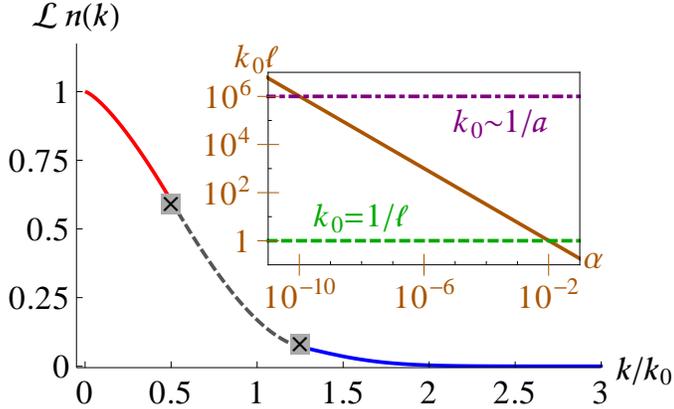}
\caption{\label{f:2} The complete wave action spectrum of Kelvin-wave turbulence at finite temperatures.  The symbol~$\mathcal{L}$ is a shorthand notation for~$\mathcal{L} = 4\pi C_{LN} \epsilon^{1/3} k^{11/3}/ \Psi^{2/3}$.  In this representation the horizontal line at~1 corresponds to the Kelvin wave spectrum at~$T=0$.  The gray line in between the crosses is an interpolation between both spectra in the crossover region~$k \approx k_0$. The inset corresponds to the estimate~\eqref{est1} for~$k_0$. }
\end{figure}

The conclusion of this part is that the energy spectrum of Kelvin-waves in the presence of dissipation in the differential approximation~(\ref{e-dif}) can be found with excellent accuracy.  Moreover, very reasonable accuracy is achieved at the first step (with the function $f_0$), that can be written in dimensional form as follows:
\begin{subequations}\label{res}
\begin{equation}\label{resA}
E(k) = C_{_{\rm LN}}\frac{\Lambda\, \kappa \,  \epsilon_0^{1/3}}{\Psi^{2/3} \, k^{5/3}} \Big[1 - \Big( \frac{k}{k_{0}} \Big)^{4/3} \Big]\,,
\end{equation} where the position of the final point $k_0$ is given by
\begin{equation}\label{resB}
k_{0}=\Big (\frac 83 \Big )^{3/4}k_{\star}=  \frac{A \,\sqrt{\Psi \epsilon_0}}
{\big(\sqrt{\alpha C_{_{\rm LN}}}\Lambda\kappa\big )^{3/2}}\,,
\end{equation}
where $A\equiv \Big ( 32 \pi / 3 \Big )^{3/4}\approx 14$. This predicted spectrum is compared to the spectrum at~$T=0$ in Fig~\ref{f:2}.   Note that this is a powerful prediction since all of the temperature dependence via the parameter~$\alpha$ is absorbed in~$k_{0}$.  This means that the energy spectrum, as well as its corresponding energy flux
\begin{equation}\label{resC}
\epsilon(k) = \frac{\epsilon_0}{5} \Big [1-\Big(\frac{k}{k_{0}}\Big)^{4/3}  \Big ]^2
 \Big [5 - \Big(\frac{k}{k_{0}}\Big)^{4/3} \Big ]\,,
\end{equation}
\end{subequations}
\noindent can be expressed as universal functions of the ratio~$k/k_{0}$.

Thanks to eq.~\eqref{resB} we can estimate $k_0$ by relating the 3D-rate of energy (per unit mass) dissipation   $\varepsilon = \nu^\prime (\kappa L)^2$ to the  effective kinematic viscosity $\nu^\prime$ and to the   vortex line density $L\simeq 1/\ell^2$, see e.g.  eq.~(1) in~\cite{Golov}. Next, according to footnote $^2$ we connect $\varepsilon\simeq \epsilon_0/\ell^2$ and get finally:
 \begin{equation}\label{est1}
k_0\ell \simeq \frac {A \, \sqrt { (\nu^\prime/\kappa) \Psi}}{\big [ \Lambda \sqrt {\alpha C\Sb {LN}}\big ]^{3/2}}\simeq \frac {0.03}{\alpha^{3/4}}\ .
\end{equation}
 In  the ``Manchester spin-down" experiment in $^4$He~\cite{Golov} it was found that for $T$ below 0.5 K (i.e. for $T/T_\lambda < 0.25$) $\nu^\prime\simeq 3 \cdot 10^{-3}\kappa $. Taking for concreteness  $\Psi\simeq 1$ and $\Lambda\simeq 15$ we get the numerical estimate given in eq.~\eqref{est1} and plotted in the inset of Fig.~\ref{f:2}.  We conclude that for $T \simeq 1$K ($\alpha \simeq 10^{-2}$), we have $k_0\ell\simeq 1$ meaning that the Kelvin wave cascade disappears and makes way for the possibility of thermally exited waves. According to the estimate~\eqref{est1} the cascade reaches the vortex-core scale~$1/a$ for very low temperatures $T\sim 0.07\,$K  ($\alpha\sim 10^{-10}$).  For even lower temperatures, the energy flux should predominantly convert into the production of quasi-particles (such as phonons) which may be observed experimentally.

\section{Reaching the edge of the window of locality}

It becomes important at this point to remember that the original collision integral eq.~(\ref{KEc}) is convergent under the assumption that the wave action spectrum behaves as a power law whose exponent is constrained to the following interval
$
n(k) \propto k^{-x} \, ; \quad 2 < x < 9/2 
$.
When~$T=0$, the exponent~$x=11/3$ clearly falls within the range of this window of locality.  However our new corrected spectrum in the quasi-inertial range for~$T\neq0$ is expected to display an increase of the local slope (that can be considered as a ``current" exponent) as one gets closer and closer to~$k_0$.  This suggests that there must exist a~$k\sb{non-loc}< k_0$ at which the wave action spectrum no longer satisfies the constraint of locality.  This maximal wavevector~$k\sb{non-loc}$ can be determined by approximating the new wave action spectrum as a local power law with a $k$-dependent local slope which is given by the logarithmic derivative
\begin{equation}
x = \lim_{k\rightarrow \psi k_0} \frac{\partial \ln n(k)}{\partial \ln k} = \frac{11}{3} \left( 1 + \frac{4}{11} \, \frac{\psi^{4/3}}{1-\psi^{4/3}} \right) \, .
\end{equation}
This allows us to demonstrate that locality will be lost when~$x\sb{non-loc} = 9/2$ which immediately translates into
 $ k\sb{non-loc} \approx  {k_0}/{2}$.
This result means that our new wave action spectrum can only be realized physically up to wavevectors about half as the predicted terminal wavevector~$k_0$.  In the next section we show how to go beyond this level of description by deriving an extension of the wave action spectrum for wavevectors that are much larger than~$k_0$.  The intermediate regime of wavevectors of the order of~$k_0$ is left for discussion in the conclusion.

\section{Energy spectrum in the far-dissipation range}

The differential approximation was based on the assumption that the energy spectrum is at least twice differentiable function of $k$. Nevertheless it lead to the spectrum~\eqref{res} in which the second derivative does not exists at $k=k_0$. This means that in the vicinity of $k_0$ this approximation is not self-consistent. This is the reason that it leads to the un-physical result that $E(k)<0$ for $k>k_0$.

In order to find the correct behavior of $E(k)$ in the dissipative region $k>k_0$ we should return to the full formulation of the problem~\eqref{EB} with the explicit form~\eqref{KEc} of the collision integral. To simplify the analysis we analyze eq.~(\ref{KEc}) in the far-dissipative region, where $k\gg k_0$.   In that region it is natural to assume that $n(k) \ll n(k_{0})$.  We will see that this ``bootstrap'' assumption is justified since~$n(k)$ decays exponentially fast.  The expansion of~$\mbox{St}_k$ turns out to be a laborious undertaking that requires a degree of vigilance; the logical steps are explained in the Appendix.  We find that the leading term when both these limits are taken into account in the form
\begin{equation}
\mbox{St}_k = - 81.6 \, \alpha C\Sb {LN}^3 \, \omega_{k_0} \, \frac{\partial^2 n_k}{\partial k^2} \, k_0^2 \int_0^1 \varphi^{8/3} \left( 1 - \varphi^{4/3} \right)^2 \upd \varphi \, ,
\label{resExp}
\end{equation}
where~$\varphi$ is a dummy variable of integration.  The collision integral must now be balanced by the dissipation due to the mutual friction.  This leads us to the following equation for the wave action spectrum~$n(k)$
\begin{equation}
\mbox{St}_k = -  \alpha \Lambda \kappa  k^2 n_k/ 4\pi\ .
\label{farDissip}
\end{equation}

\noindent Restricting ourselves only to decaying solutions and the limit of large~$k$, it is not difficult to see that the general solution of~eq.~(\ref{farDissip}) can be simplified into
\begin{subequations}\label{far-dis} \begin{eqnarray}\label{far-disA}
n_k &=&  c_1 \, \frac{4\pi C\Sb {LN} \epsilon_0^{1/3}}{\Psi^{2/3} k_{0}^{11/3}} \sqrt{\frac{\tilde{k}}{k}} \exp \left( - \frac{k^2}{4 \tilde{k}^2} \right) \,, \ \mbox{where}~~~~ \\ \label{far-disB}
\tilde{k} &\approx&  \frac{25}{2\alpha^{3/4}} \sqrt{\frac{\epsilon_0 \Psi}{\kappa^3 \Lambda^3}} \approx 0.37 \, k_{0}\,,
\end{eqnarray}
\end{subequations}
\noindent Note that even though this result is presented here as an expansion of the ``effective'' 4-wave~$(1\Leftrightarrow 3)$ collision integral, eq.~(\ref{KEc}), it does in fact remain identical in the more fundamental context of the original 6-wave~($3\Leftrightarrow 3$) collision integral of~\cite{rudenko}.  We are now in possession of~2 different expressions for~$n(k)$.  However, since the corrected wave action spectrum presented in the previous section is only valid for~$k\ll k_{0}$ and the one presented in this section is the result of an expansion of the collision integral for~$k \gg k_{0}$, there is no reason to expect that these two solutions should be matched in their respective domain of validity.  Therefore, we choose the dimensionless constant~$c_1 = 1$.  Nevertheless the crossover region between these two different regimes is very narrow around the wavevector $k \approx k_{0}$ as one can see from Fig~\ref{f:2}.

\section{Conclusion}

The main message of this letter is that the presence of a small but finite temperature is enough to dramatically suppress the Kelvin-wave turbulence from the simple scale-free~eq.~(\ref{LN}) to the complex behavior summarized in Fig.\ref{f:2}.

Since our theoretical analysis is limited in the crossover region, we believe that important additional information can be gained from numerical simulations of finite temperature Kelvin-wave turbulence. However preliminary attempts in this direction~\footnote{Preliminary numerical simulations~\cite{JN} of weak wave turbulence reveal that finite size effects play in an important role which is further amplified in this one-dimensional setting. This  inhibits wave resonances and leads to appearance of  several
asymptotic ranges. This results in great difficulties in choosing the parameters of the simulations.}  show that this is a tricky task (even for a 1D
systems like ours) which would require significant efforts in the future.

The finite temperature results presented in this letter appear to constitute the first attempt to bring the theoretical predictions of Kelvin-wave turbulence closer to realistic situations. Hopefully these can be used in actual experiments on superfluid turbulence. There are however still many questions that lie ahead.  For example, how does the underlying assumption of weak turbulence, which ignores the all-important possibility of quantum vortex reconnections, affect the energy spectrum?  How to go beyond this approximation?

\section{Appendix: Expansion of collision integral}

The convergence of the collision integral is ensured only through a series of deep cancellations that deserve to be explained in more detail.  As mentioned in the text, we are interested in the limit where one of the wavevector~$k$ is much larger than the terminal point~$k_0$.  In addition, we make the assumption that~$n(k) \ll n(k_0)$.  First, let us define the small parameter~$\mu = k_0 / k \ll 1$ and for let us also refer to the conservation laws of the first line in the integral~\eqref{KEc} as ``spherical'' and the conservation laws of the second line as ``hyperbolic''.   In this case, it is clear that the hyperbolic geometry requires that~$k_1$ be close to~$k$ while~$k_2$ and~$k_3$ should be much smaller, of the order of~$k_0$.  On the other hand, the spherical geometry is more permissive and allows for any of the~3 wavevectors ($k_1$, $k_2$ or~$k_3$) to be close to~$k$.  For example, if it is~$k_1$ which is of order~$k$ then it is~$k_2$ and~$k_3$ which will be of order~$k_0$.  In this case we can write
\begin{equation}
{\bm k}_2 = \phi {\bm k}_{0}\,\,; \quad -1 \leq \phi \leq 1. \tag{A.1}
\label{appEq}
\end{equation}

\noindent Because of the other~2 possibilities, it is now clear that the expansion of the part of the collision integral constrained by the spherical geometry should be multiplied by a factor of~$3$.  Notice that this new prefactor matches exactly the prefactor of the hyperbolic geometry albeit with a different sign.  This feature has already been discussed in~\cite{victor1} and is responsible for cancellations of all the terms of order~$\mu^2$,~$\mu^4$ and~$\mu^6$.  Since some of these terms would otherwise be non-integrable these cancellations are quite outstanding.  For the sake of the discussion, let us now go back the special case (spherical geometry) mentioned above in~\eqref{appEq}.  First of all one should notice that because of the~$\{ {\bm k}_2 , - {\bm k}_2 \}$ symmetry, all of the odd-ordered terms of the expansion identically vanish because of the integration over~$\phi$.  Interestingly it turns out that all of the terms of order~$\mu^8$ involve quadratic combinations~$n(k)^2$ which must be much smaller than~$n(k)$ according to our ``bootstrap'' assumption stated above.  Since the next terms of order~$\mu^9$ are absent, it is necessary to study terms of order~$\mu^{10}$.  Quite remarkably, we discovered that there is only one term of this order that is linear in~$n(k)$; It involves the second derivative of the wave action spectrum and is presented in~eq.~\eqref{resExp}

\acknowledgments
We acknowledge S. Nazarenko and J. Laurie for useful discussions and sharing their preliminary results of numerical simulations of Kelvin-wave turbulence before publication.  We are also grateful to both anonymous referees for numerous critical comments and suggestions which helped us to greatly improve this letter.
This work is supported by the EU FP7 Microkelvin program  (Project No.~228464) and by  the U.S. Israel Binational Science Foundation.

\end{document}